\documentclass[a4paper,11pt]{article}

\usepackage[a4paper,margin=2.5cm]{geometry}

\usepackage[utf8]{inputenc}

\usepackage[english]{babel}

\usepackage{amsfonts}

\usepackage{graphicx}

\usepackage{subfigure}

\usepackage[thinlines]{easytable}

\usepackage{natbib}

\setcitestyle{authoryear,round,semicolon}

\usepackage{hyperref}

\usepackage{algorithm}
\usepackage{algpseudocode}

\title{The Economic Value of Depth}

\author{Pedro Afonso Fernandes\footnote{ORCID: \url{https://orcid.org/0000-0001-5762-5157}. Correspondence: Universidade Católica Portuguesa, Católica Lisbon School of Business \& Economics, Palma de Cima, Building 5, 4th floor, Room 5430, 1649-023 Lisboa, Portugal. Email: paf@ucp.pt. URL: \url{https://paf.com.pt}.}\\\\Universidade Católica Portuguesa\\Católica Lisbon School of Business \& Economics\\Católica Lisbon Research Unit in Business \& Economics (CUBE)\\Portugal}

\date{\selectlanguage{english} \today}

\begin{document}

\maketitle

\selectlanguage{english}

\begin{abstract}

The main goal of this article is to introduce an economic perspective in the social logic of space. Firstly, we describe the economic model of a linear city to show how depth can generate value by creating local monopolies in less integrated spaces. Then, a new syntactic measure, the d-value, is proposed to capture the relation between the depth of some space from outside and the mean depth of all spaces from outside. An application to a public housing estate suggests that economic activities and services may be located in spaces with a d-value close to one. The article is complemented by a Prolog programme with a special predicate to compute the d-value.\\\


\end{abstract}

\pagebreak


\pagebreak

\section{Introduction}
\label{sec:intro}

Space syntax is a set of techniques for analysing urban settlements or buildings and theories linking space and society founded on architecture, engineering, mathematics, sociology, anthropology, ethnography, linguistics, psychology, biology, and computer science. It was developed originally by Bill Hillier, Julienne Hanson, and colleagues at the Bartlett School of Architecture and Planning, University College of London (UCL) since the 1970s. Their innovative approach was condensed into three landmark books: The Social Logic of Space \citep{Hillier1984}, Decoding Homes and Houses \citep{Hanson1998}, and Space is the machine: a configurational theory of architecture \citep{Hillier2007}.

Space syntax is mainly concerned with the study of the relations between convex spaces. Convexity exists when straight lines can be drawn from any point in a space to any other point in it without leaving the space itself \citep{Hillier1984}. In fact, convex spaces rather than concave spaces stimulate social interaction in the sense that everyone sees everyone within that kind of space. Thus, the starting point of a syntactic analysis is typically a map of the fattest convex and open (or permeable) spaces that cover the settlement (or building) in question. Then, several syntactic measures can be computed to explore the connectivity and asymmetry between convex spaces or the minimum set of axial lines that covered them.

In this scope, the notion of depth is concerned with the number of intervening spaces through which it is necessary to go to get from one space to another \citep{Hillier1984}. The deepest or asymmetric spaces are typically located at the end or limit of a complex because we may have to take several steps to reach the other spaces. In contrast, integrated spaces are typically located in the 'middle' of the settlement. Thus, depth has an economic meaning in the sense that it represents the cost of moving between spaces. In fact, the most valuable spaces typically have a small mean depth, that is, we do not need to take too much steps from them to get the other spaces in the complex. 

The economic meaning of integration can be found in the high rents typically paid in the central business district (CBD) of most towns, or in neighbourhoods well connected with highways or railways. However, depth itself can have an economic value. In addition to the value of peace and quiet, suburban consumers may be willing to pay higher prices for goods sold near their homes to avoid freight or commuting costs or because they like local stores. This kind of value is not captured by typical syntactic measures.

In this article, we describe the economic model of a linear city to show how depth can generate value by creating local monopolies in less integrated spaces. Then, a new syntactic measure is proposed, the d-value, to capture the economic value of depth. An application to a public housing estate located in Odivelas, Portugal, used Prolog, a logic programming language developed in the 1970s \citep{Colmerauer1993}, to illustrate how the d-value can be easily computed. 

Prolog applies logic to represent knowledge and uses deduction to solve problems by deriving logical consequences \citep{Kowalski1988}. A corollary of using logic to represent knowledge is that such knowledge can be understood declaratively. This kind of reasoning is embodied in space syntax when it represents spatial arrangements as a field of knowables, that is, as a system of possibilities governed by a simple and underlying system of concepts \citep{Hillier1984}.  A comprehensive description of the application of logic programming to space syntax can be found in \citet{Fernandes2023}.

The material in this article is complemented by an open source computer programme stored in SWISH (\url{https://swish.swi-prolog.org/p/gulbenkian.pl}, accessed on 12 June 2025). SWISH is the online version for the sharing of SWI-Prolog, a free and versatile implementation of the Prolog language developed by Jan Wielemaker and colleagues at the University of Amsterdam, The Netherlands, since 1987~\citep{Wielemaker2012}.

\section{Basic Concepts}
\label{sec:framework}

\begin{figure}[h]
\centering
\includegraphics[width=14cm]{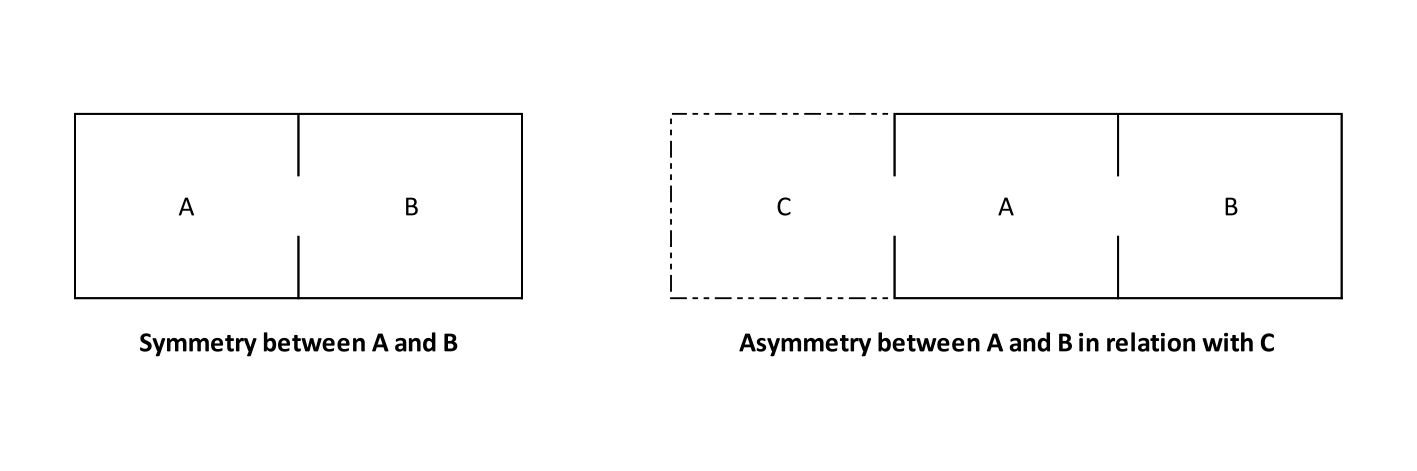}
\caption{Basic configurations.}
\label{fig:basic_config}
\end{figure}

Based on graph theory and computer-aided simulations, space syntax aims to find and explain the relation between spatial configurations and social activities. \emph{Configuration} is a concept that addresses the whole of a complex (settlement or building) rather than its parts and captures how the relations between two convex spaces, say A and B, might be affected by a third space C \citep{Hillier2007}. For example, if A and B are adjacent or permeable, then they have a symmetric configuration in the sense that, if A is the neighbour of B, then B is the neighbour of A, as illustrated on the left-hand side of Figure~\ref{fig:basic_config}.

However, if only A is connected with a third space C, as in the right-hand side of the same figure, A and B become asymmetrical in relation to C because we have to pass through A to get to B from C, but we do not have to pass through B to get to A from C. Thus, \emph{asymmetry} relates to \emph{depth}, that is, with the number of spaces (or steps) necessary to go from a certain space, say C, to another space, A (1 step), B (2 steps), and so on.

\begin{displaymath}
C \rightarrow A \rightarrow B \rightarrow \ldots
\end{displaymath}

If we count the number of steps necessary to go from a certain space to every other space in a complex, we can obtain a measure of its total depth (TD), or mean depth (MD) by dividing that total by the number of spaces in the complex minus one, the original space \citep{Hillier1984} \citep{Hanson1998}. In the previous example, the total depth of C is 3, which is the sum of steps to reach A and B (1 + 2) from that origin. This is also the case for B, but the total depth of A is 2, noting that this (central) space is directly connected to the spaces B or C (1 + 1). Thus, the mean depth is 1 (2/2) for A and 1.5 (3/2) for B or C, noting that the number of spaces minus one in this simple complex is 2 (3 $-$ 1).

This simple illustration suggests that syntactic measures can be computed for every convex space in a settlement (or building). Then, we may eventually find that some spaces have a lower depth than all other spaces (A in the same example), and others have a greater depth (B and C). The former are the most integrated spaces, where social life may be concentrated in cities \citep{Heitor2015}. The latter are typically the most segregated, quiet, or remote spaces in a complex. Thus, \emph{integration} is inversely related to depth. It is a global measure in the sense that it considers the configuration of a certain space in relation to all other spaces. In addition, local measures such as control are based on the relations between each space and only the spaces directly connected to it \citep{Turner2004}.

Traditionally, high spatial integration of the street network is a sufficient condition for economic centrality \citep{vanNes2022}. Thus, the places where trade, shopping, and finances take place may be the most integrated streets on a local or global scale. Accessibility, visibility, permeability, and connectivity to direct side streets are also necessary conditions for the concentration of stores. However, as the following model suggests, depth can protect local markets, especially with high transportation costs. Thus, asymmetry can favour profits.

\section{Model}
\label{sec:model}

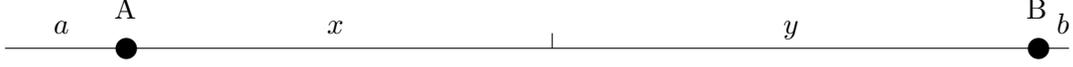
\begin{figure}[h]
	\centering
	\setlength{\unitlength}{4mm}
	
	\begin{picture}(35,3)
		\put(0,0){\line(1,0){35}}
        \put(1.6,0.5){\normalsize$a$}
        \put(3.6,1){\normalsize{A}}
        \put(4,0){\circle*{0.7}}
		  \put(10.6,0.5){\normalsize$x$}
		\put(18,0){\line(0,1){0.5}}
        \put(25.6,0.5){\normalsize$y$}
		\put(33.6,1){\normalsize{B}}
        \put(34,0){\circle*{0.7}}
		  \put(34.6,0.5){\normalsize$b$}
	\end{picture}
	
	\caption{Hotelling's linear city.}
	\label{fig:linear_city}
\end{figure}

The linear city economic model was originally developed by Harold \citet{Hotelling1929} to deal with the instability of the duopoly model, where a small change in the price of either company takes away all the opponent's business. In Hotelling's Main Street, the buyers of a homogeneous good are uniformly distributed along a line of length $l$. At distances $a$ and $b$ respectively from the two ends of this line are stores A and B (Figure \ref{fig:linear_city}).

Let us denote $p_1$ the price of A, $p_2$ the price of B, $q_1$ the quantity sold by A and $q_2$ the quantity sold by B. Each consumer is indifferent between buying from A or B except in terms of price plus transportation cost between the store A or B and her home at a cost $c$ per unit of distance. If store A wants to sell something, its price $p_1$ should be below $p_2$ plus the transportation cost from store B, that is, $p_1 < p_2 + c(l-a-b)$. In this case, store A sells the quantity $a$ plus the quantity $x$ which depends on the difference between prices $p_1$ and $p_2$.  Similarly, store B sells $b$ plus $y$ if $p_2 < p_1 + c(l-a-b)$. Once again, $y$ is a function of the difference between prices. 

Thus, stores are monopolists at the respective end of the line: A gets the demand $a$, and B the demand $b$. Between locations A and B, stores have to split the market based on their (optimal) prices. The point of division between the regions served by A and B, respectively $x$ and $y$, is determined by the condition of indifference to buy from one store or another, that is,  

\begin{equation}
\label{eq:indifference}
p_1 + c x = p_2 + c y.
\end{equation}

\noindent Noting that $a + x + y + b = l$, \citet{Hotelling1929} found from (\ref{eq:indifference}) that

\begin{equation}
\label{eq:x}
x = \frac{1}{2} \left( l - a - b + \frac{p_2 - p_1}{c} \right), \textrm{ and}
\end{equation}

\begin{equation}
\label{eq:y}
y = \frac{1}{2} \left( l - a - b + \frac{p_1 - p_2}{c} \right).
\end{equation}

Thus, the quantity $x$ sold by store A increases with the price of the other store $p_2$ or when its own price $p_1$ decreases. Similarly, store B sold more when $p_1$ increased or $p_2$ decreased. Obviously, the quantity to be split $x+y$ decreases with length $a+b$. Having some end of the line minimises the competition in the middle of the line.

The optimal price for each store can be found by equating the first derivative of its own profit to zero, the so-called first-order condition. Noting that the profits of both firms are 

\begin{equation}
\label{eq:profit_1}
\pi_1 = p_1 q_1 = p_1 (a+x) = \frac{1}{2} \left( l + a - b \right) p_1 - \frac{{p_1}^{2}}{2c} + \frac{p_1 p_2}{2c}, \textrm{ and}
\end{equation}

\begin{equation}
\label{eq:profit_2}
\pi_2 = p_2 q_2 = p_2 (b+y) = \frac{1}{2} \left( l - a + b \right) p_2 - \frac{{p_2}^{2}}{2c} + \frac{p_1 p_2}{2c}, 
\end{equation}

\noindent from the first-order conditions described by \citet{Hotelling1929}, we obtain the optimal prices

\begin{equation}
\label{eq:x_price}
p_1 = c \left( l + \frac{a - b}{3} \right), \textrm{ and}
\end{equation}

\begin{equation}
\label{eq:y_price}
p_2 = c \left( l + \frac{b - a}{3} \right).
\end{equation}

Thus, if a store is located far away from the end of the city and if it does not face local competition until that end, which requires some depth from the middle of the line, it should adopt a higher price. In fact, the optimal price of store A ($p_1$) increases with length $a$ and the same applies to store B, that is, $p_2$ increases with $b$. Similarly, store A or B can achieve higher sales and profits from length $a$ or $b$, respectively, noting that optimal quantities are 

\begin{equation}
\label{eq:x_quantity}
q_1 = a + x = \frac{1}{2} \left( l + \frac{a - b}{3} \right), \textrm{ and}
\end{equation}

\begin{equation}
\label{eq:y_quantity}
q_2 = b + y = \frac{1}{2} \left( l + \frac{b - a}{3} \right).
\end{equation}

A small example can illustrate why $a > b$ implies higher prices, quantities, and profits for store A. Following \citet{Hotelling1929}, let $l = 35$, $a = 4$ and $b = 1$ as in figure \ref{fig:linear_city}, with $c = 1$. Then, from equations (\ref{eq:x_price}) and (\ref{eq:y_price}), we get $p_1 = 35 + (4 - 1)/3 = 36$ and $p_2 = 35 + (1 - 4)/3 = 34$. Similarly, from (\ref{eq:x_quantity}) and (\ref{eq:y_quantity}), we get $q_1 = (35 + 3/3)/2 = 18$ and $q_2 = (35 - 3/3)/2 = 17$. Thus, the profit of store A will be $36 \times 18 = 648$ euros, while store B will get $34 \times 17 = 578$ euros.     

If the location of B is fixed, A has an economic incentive to make $a$ as large as possible. This means that it will come as close to B. However, if A and B are too close, the system becomes unstable like the classic duopoly: each store can get the full market by slightly decreasing its own price. Thus, A and B should maintain some distance between them. As \citet{Hotelling1929} said: "The intermediate segment of the market acts as a cushion as well as a bone of contention; when it disappears we have Cournot's case, and Bertrand's objection applies".

\section{Implications}
\label{sec:implications}

What are the implications of Hotelling's model on space syntax? Despite its simplicity and theoretic nature, Hotelling's model suggests that depth can matter in the sense that a less integrated space can provide local demand and less competition for firms located there. Central or integrated spaces are typically very tough: In addition to high rents, companies may have to lower their prices to sell their goods and services. In contrast, peripheral locations can offer higher profits. However, companies located in very deep locations may only get a small portion of the total market, such as firm B in Figure \ref{fig:linear_city} which gets $b << a$. Thus, a good location may be an intermediate point between the middle and end of the settlement, keeping some intermediate space between stores.

This evidence suggests a new syntactic measure, d-value, that captures the relation between the depth of some space from outside ($D_o$) and the mean depth of all spaces from outside ($MD_o$), that is,

\begin{equation}
\label{eq:d_value}
\textrm{d-value} = \frac{D_o}{MD_o}.
\end{equation}

In this framework, a d-value close to 0 means that the space is too close to the end of the settlement, where the market dimension may be tiny. In contrast, a d-value greater than 1 means that the space is located close to the centre of the settlement, where competition may be hard. Ideally, the space should have a d-value close to 1, which means that its depth from outside is roughly equal to the mean depth from outside.

\section{Application}
\label{sec:application}

\begin{figure}[!ht]
	\centering
	\includegraphics[width=13.1cm]{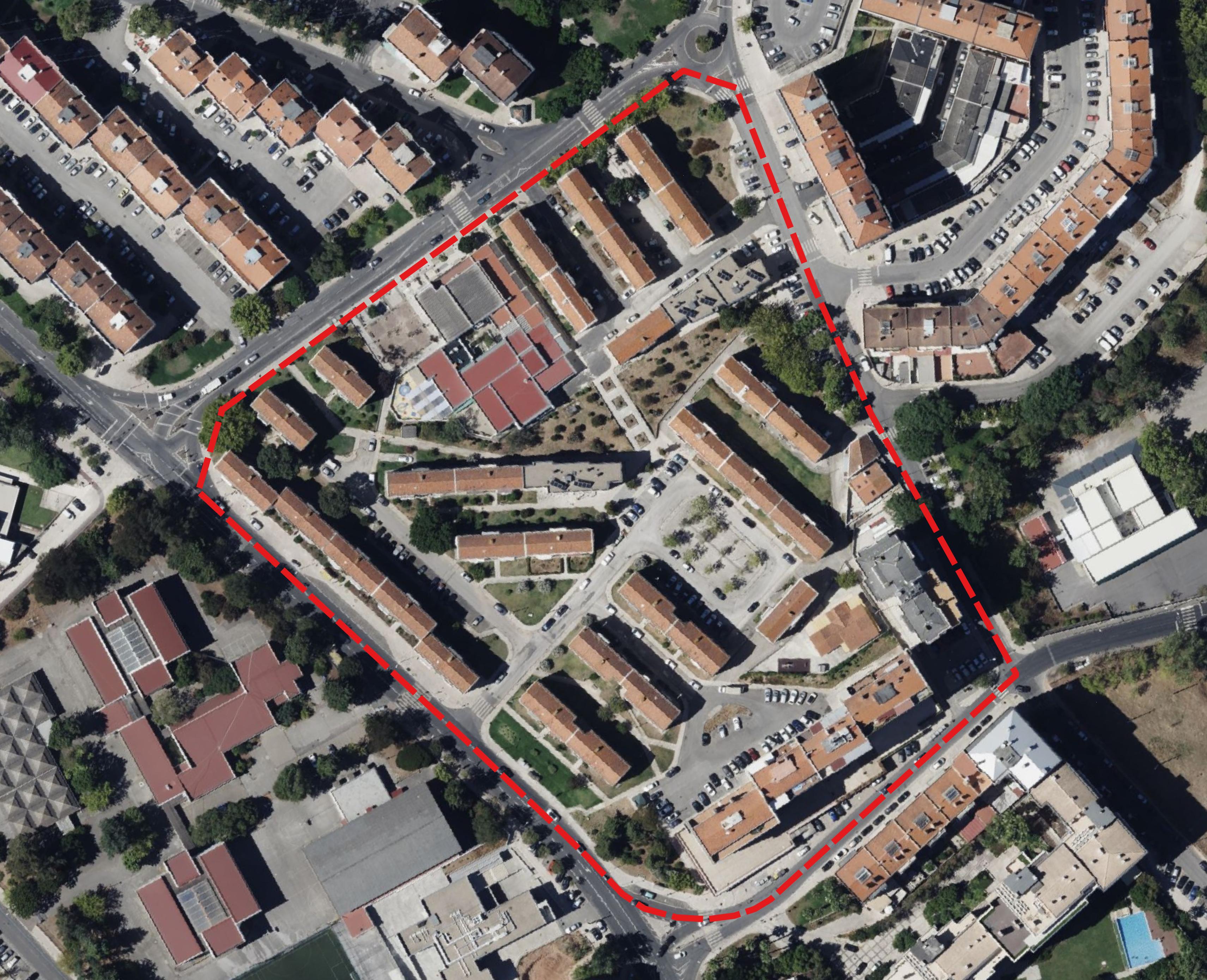}
	\caption{Bairro Gulbenkian: aerial photograph.}
	\label{fig:orto}
\end{figure}

Bairro Gulbenkian (figure \ref{fig:orto}) is a public housing estate built in the late 1960s in Odivelas, near Lisbon, Portugal. It occupies a plot of 28,500 square metres and is the largest estate built with the financial and technical support of the Calouste Gulbenkian Foundation in a special resettlement plan to deal with the homeless of the great flood of November 25-26, 1967, in the Great Lisbon, which killed about 700 people \citep{Malheiros2018}. Composed of 15 isolated blocks with three storeys and a total of 160 dwellings (about 600 inhabitants), Bairro Gulbenkian is a good case for testing and developing new syntactic methods. 

\begin{figure}[!ht]
	\centering
	\includegraphics[width=13.1cm]{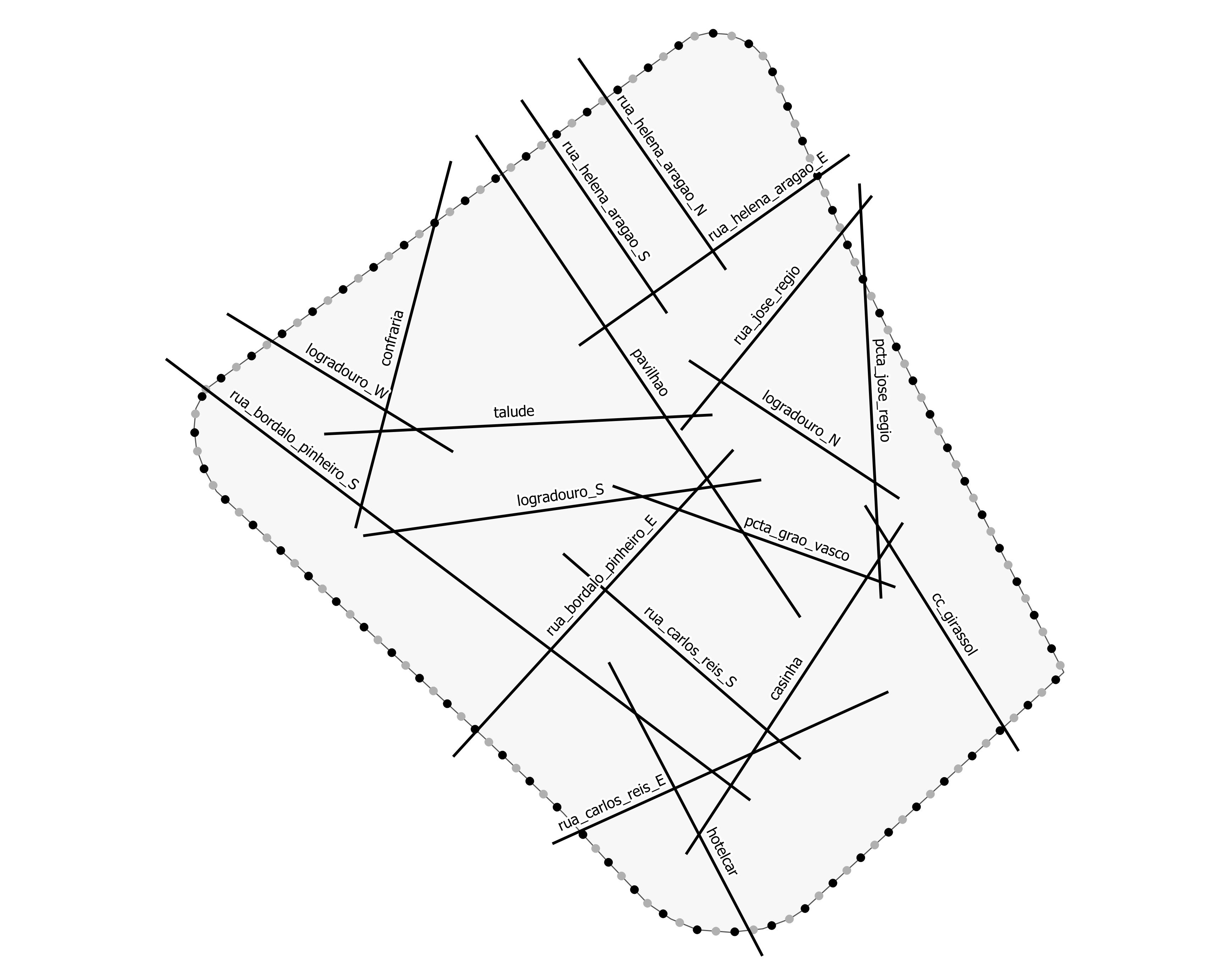}
	\caption{Bairro Gulbenkian: axial map (dotted line: outside).}
	\label{fig:axial}
\end{figure}

The axial map of Bairro Gulbenkian (figure \ref{fig:axial}) reveals that the maximum depth from the outside is 2, that is, the lines that cover its convex spaces are one or two steps away from the end of the settlement.

The mean depth from the outside ($MD_o$) is 1.32. It can be easily computed with the Prolog programme available at \url{https://swish.swi-prolog.org/p/gulbenkian.pl} (accessed on 12 June 2025) by asking the following query:

\begin{verbatim}
?-meandepth(outside,Y,D,MD).
\end{verbatim}

\noindent This computation is possible because we declared all the lines connected with 'outside' at the beginning of the programme, namely,

\begin{verbatim}
connected(outside,hotelcar,1).
\end{verbatim}

Thus, axial lines with depth 1 from the outside, such as 'pavilhao' or 'hotelcar', have a d-value of $0.76 = 1/1.32$. Similarly, axial lines with depth 2, such as 'talude' or 'pcta\_grao\_vasco', have a d-value of $1.52 = 2/1.32$. These figures can be computed with the following Prolog query:  

\begin{verbatim}
?-dvalue(X,Y,Z,D,Q).
\end{verbatim}

\noindent that uses the predicate \verb!dvalue! defined in the main programme as

\begin{verbatim}
dvalue(X,Y,Z,D,Q):-depth(outside,X,D), meandepth(outside,Y,Z,MD), Q is D/MD.
\end{verbatim}

Interestingly, most stores and services are located in spaces with depth 1, that is, with d-value close to one. This happens with the sports hall, nursery, elderly centre, and coffee shop, all located at 'pavilhao', the mall, located at 'cc\_girassol', and the garage, located at 'hotelar'.

\section{Conclusion}
\label{sec:conclusion}

The Hotelling model is founded on several unrealistic assumptions, namely, that the market is a finite line where consumers are uniformly distributed and take decisions only on the basis of price plus transportation costs linear to distance, there are no production costs or there is only one identical good sold on the market by two firms, an assumption that can be easily relaxed \citep{Economides1993}. However, this model describes the essence of economic decision within a spatial framework: the distance between consumers and producers matters because it imposes a cost, protecting local markets. Thus, a store located at some depth may have some market power, noting that optimal profits increase with the unit cost of transportation $c$:

\begin{equation}
\label{eq:x_profit}
\pi_1 = p_1 q_1 = \frac{c}{2} \left( l + \frac{a - b}{3} \right)^{2}, \textrm{ and}
\end{equation}

\begin{equation}
\label{eq:y_profit}
\pi_2 = p_2 q_2 = \frac{c}{2} \left( l + \frac{b - a}{3} \right)^{2}.
\end{equation}

Thus, firms can be favoured locally when transport is difficult for consumers \citep{Hotelling1929}. Because profits also depend on the distance to the end of the line, $a$ or $b$, stores should be located in an intermediate point between that end and the centre of the city. The proposed d-value captures this idea by comparing the depth from outside with the average depth of all possible locations. With this simple syntactic measure, we can introduce some economic reasoning to the social logic of space.

\section*{Acknowledgments}
\label{sec:acknowledgements}

This work was financed by Fundação para a Ciência e a Tecnologia (FCT) under a doctorate auxiliary researcher grant from Universidade Católica Portuguesa (UCP) - Católica Lisbon Research Unit in Business \& Economics (CUBE) with the digital object identifier (DOI): \href{https://doi.org/10.54499/CEECINST/00070/2021/CP1778/CT0008}{10.54499/CEECINST/00070/2021/CP1778/CT0008}.

\bibliography{library}

\end{document}